\documentclass[english]{article}
\usepackage[T1]{fontenc}
\usepackage[latin9]{inputenc}
\usepackage[letterpaper]{geometry}

\usepackage{amsmath}
\usepackage{amssymb}
\usepackage{graphicx}
\usepackage{esint}
\usepackage{lineno}
\usepackage[authoryear]{natbib}
\usepackage{setspace}

\makeatletter

\newcommand{\lyxaddress}[1]{
\par {\raggedright #1
\vspace{1.4em}
\noindent\par}
}

\date{March 8, 2017}
\usepackage{chngpage}

\makeatother

\usepackage{babel}
 
\begin{document}
	{
	\setstretch{1.0}
	\title{Local equilibrium and retardation revisited\footnote{Los Alamos National Laboratory technical report: \textbf{LA-UR-16-22097}}}

	\author{Scott K. Hansen and Velimir V. Vesselinov}
	
	\maketitle
	
	\lyxaddress{Earth and Environmental Sciences Division, Los Alamos National Laboratory,
	Los Alamos, NM 87545, USA.}
	
	\begin{abstract}
		In modeling solute transport with mobile-immobile mass transfer (MIMT), it is common to use an advection-dispersion equation (ADE) with a retardation factor, or \textit{retarded ADE}. This is commonly referred to as making the \textit{local equilibrium assumption}. Assuming local equilibrium (LE), Eulerian textbook treatments derive the retarded ADE, ostensibly exactly. However, other authors have presented rigorous mathematical derivations of the dispersive effect of mass transfer, applicable even in the case of arbitrarily fast mass transfer. First, we resolve the apparent contradiction between these seemingly exact derivations by adopting a Lagrangian point of view. We show that LE constrains the expected time immobile, whereas the retarded ADE actually embeds a stronger, nonphysical, constraint: that all particles spend the same amount of every time increment immobile. Eulerian derivations of the retarded ADE thus silently commit the gambler's fallacy, leading them to ignore dispersion due to mass transfer that is correctly modeled by other approaches. Second, we present a numerical particle tracking study of transport in a heterogeneous aquifer subject to first-order MIMT. Transport is modeled (a) exactly, and then (b) approximated with the retarded ADE. Strikingly different results are obtained, even though quasi-LE is maintained at all times by the exact MIMT simulation. We thus observe that use of the phrase \textit{local equilibrium assumption} to refer to ADE validity is not correct. We highlight that solute remobilization rate is the true control on retarded ADE validity, and note that classic "local equilibrium assumption" (i.e., ADE validity) criteria actually test for insignificance of MIMT-driven dispersion relative to hydrodynamic dispersion, rather than for local equilibrium. 
	\end{abstract}
	}
\pagebreak

\section{Introduction}

	This work offers some observations on a classic topic: the relationship between mobile-immobile solute equilibrium, kinetics, and the retarded advection-dispersion equation (ADE).
	
	Our primary motivation for revisiting this subject lies in the fact that two contradictory approaches to modeling mobile-immobile mass transfer have co-existed in the literature for decades, both underpinned by seemingly exact mathematical arguments. The first approach---usage of a retardation factor in the ADE---has been ostensibly derived under fast mass transfer, or "local equilibrium" conditions. In parallel, a second group of authors have shown that mass transfer is always dispersive and, for first-order kinetic mass transfer, analytically quantified its effect.	We concur with this second group of authors that mass-transfer is always dispersive. However, the error underpinning the ostensibly exact retarded ADE derivation does not appear to have been pinpointed in the literature. Furthermore, we document below numerous places in the literature where the retarded ADE is treated as exact, in which this error is not harmless. 
	
	Our secondary motivation is to correct a potential misconception regarding the relationship between degree of local equilibrium and degree of validity of the retarded ADE. While usage of the latter is sometimes referred to as the "local equilibrium assumption" (LEA), we present an example transport simulation that respects local equilibrium, but in which the retarded ADE is a very poor proxy for true behavior. It is actually the case that the solute remobilization rate is the control on the validity of the retarded ADE.
	
	The remediation of contaminated groundwater sites is a topic of persistent interest in industrialized societies. Remediation is generally quite expensive---the U.S. National Research Council recently estimated that the cost to clean up existing sites in the United States at over \$100 billion over the next 30 years \cite{us2013alternatives}. This high cost necessitates the development of accurate yet tractable groundwater models. Unfortunately, the trade-off between accuracy and tractability is made difficult by the presence of pore-scale mass transfer processes that are too complex to model at their natural support scale, but which also have macroscopically observable effects. One of the most important such processes is adsorption, the reversible chemical interaction between dissolved contaminants and solid-phase components of the pore structure. 
	
	In hydrogeologic and engineering models, it is common to model mobile-immobile (e.g., sorbing) solute transport with the retarded ADE. This is the equation: 
	\begin{equation} 
		R\frac{\partial c}{\partial t}(x,t)=-v\frac{\partial c}{\partial x}(x,t)+D\frac{\partial^{2}c}{\partial x^{2}}(x,t),\label{eq:ADE retarded}
	\end{equation} 
	where $R$ $\left[\mathrm{dimensionless}\right]$ is a constant retardation factor, $c$ $\left[\mathrm{ML^{-3}}\right]$ is aqueous solute concentration, $t$ $\left[\mathrm{T}\right]$ is time, $x$ $\left[\mathrm{L}\right]$ is the spatial coordinate, $v$ $\left[\mathrm{LT^{-1}}\right]$ is the advection velocity, and $D$ $\left[\mathrm{L^{2}T^{-1}}\right]$ is a Fickian dispersion coefficient. (In all unit expressions, $[\mathrm{M}]$ represents mass, $[\mathrm{L}]$ represents length, and $[\mathrm{T}]$ represents time.) This equation applies as well to mobile-immobile mass transfer (MIMT) processes other than sorption.
	
	The $R$ on the LHS accumulation term of \eqref{eq:ADE retarded} can be viewed in two different ways: as a rescaling factor for time, or as a rescaling factor for solute accumulation. Based on this, there are two possible understandings of the retardation factor in homogeneous media. It may be conceived as the ratio of groundwater velocity to mean solute velocity \citep[e.g.,][]{rajaram_time_1997}, or as the ratio of total (mobile and immobile) solute concentration to mobile solute concentration at equilibrium. This second conception motivates the idea that ``local equilibrium'' mass transfer provides support for usage of the retarded ADE.
	
	However, there can be no exact equilibrium under transient conditions (only fast kinetics). While this may seem innocuous, its impact may be significant. To understand the degree of approximation that is occurring relative to fast kinetic behavior, we will consider the explicit transport equations for advection and dispersion in the presence of first-order single-rate mass transfer. The relevant equations may be written \citep[][p. 133]{fetter_contaminant_1993}: 
	\begin{equation} 
		\begin{array}{rl} \frac{\partial c}{\partial t}(x,t)+\frac{\partial s}{\partial t}(x,t) & =-v\frac{\partial c}{\partial x}(x,t)+D\frac{\partial^2 c}{\partial x^2}(x,t)\\ \\ \frac{\partial s}{\partial t}(x,t) & =\lambda c(x,t)-\mu s(x,t) \end{array},\label{eq:Explicit treatment of sorption} 
	\end{equation} where $s$ $\left[\mathrm{ML^{-3}}\right]$ is the immobile concentration, $\lambda$ $\left[\mathrm{T^{-1}}\right]$ is the probability per unit time of immobilization of mobile solute, and $\mu$ $\left[\mathrm{T^{-1}}\right]$ is the probability per unit time for mobilization of immobile solute. We show in Appendix \ref{sec: eulerian derivation} how \eqref{eq:ADE retarded} is a special case of \eqref{eq:Explicit treatment of sorption}, in the $\mu \rightarrow \infty$ limit.  So while the first-order kinetic model is itself an idealization, it is no more so than the retarded ADE and additionally captures the true behavior of solute being continuously mobile or immobile for finite intervals. Equations of form \eqref{eq:Explicit treatment of sorption} are widely used in the literature to capture general MIMT processes (see conceptual discussion in \citet{Fernandez-garcia2015}, \citet{valocchi_validity_1985}, and \citet{bahr_direct_1987}). They are applicable over a range of advection velocities \citep{Zhang2008a,Zhang2009} and spatial support scales \citep{Raoof2010}. Thus, the analysis of the system they describe is relevant to a large variety of hydrogeologic problems. This single-rate paradigm, while not encompassing all forms of MIMT---for example, non-linear sorption and mobile-immobile phenomena with heavy-tailed immobile-state waiting times \citep{Margolin2003,Schumer2003} are not covered---remains of sufficient generality to reveal the nature of the retardation factor approximation.
	
	That first-order kinetic MIMT has a dispersive effect (i.e., that capture and release of particles independent of one another drives spatial spreading of the distribution of $c$) has long been recognized. In \citet{giddings_molecular_1955}, equations for the spreading of breakthrough curves at the output of a chromatograph, using essentially the assumptions of chemical non-equilibrium, were derived. \citet{valocchi_validity_1985} and \cite{Goltz1987} performed thorough parametric studies of moments for a variety of MIMT processes and their contributions to the spreading of plumes in the subsurface. Many other authors have considered aspects of this topic, as well. Regardless of the rapidity of the MIMT, the retarded ADE does not capture dispersion due to mass transfer: the time-scaling retardation factor does not change the qualitative shape of the solution by adjusting the relative weights of advection and dispersion. In fact, that \eqref{eq:ADE retarded} fails to capture dispersion encoded by \eqref{eq:Explicit treatment of sorption} was made explicitly in a numerical study by \citet{elfeki_modeling_2007}.
	
	At the same time, however, the substitution of retardation factors from equilibrium batch experiments---i.e., the use of \eqref{eq:ADE retarded}---to modify transport equations in the presence of kinetic sorption is frequently presented in expository works as though it is exact. A derivation of the retarded ADE by such means is presented as mathematically exact in the canonical \emph{Hydraulics of Groundwater} text \cite[p. 242]{bear_hydraulics_1979}, given ``equilibrium'' sorption, and in the authors' experience is believed by many hydrogeologists to be exact. \citet{bouwer_simple_1991} also developed a relationship between a soil distribution coefficient and retardation factor by assuming that all solute released at the same instant has been, at any moment, immobile for the exact same amount of time. In a recent textbook \citep[p. 208]{hiscock_kevin_hydrogeology:_2014}, the Bouwer result is also reported without any explicit indication that transport with mass transfer is a dispersive process in which different particles may be immobile for different fractions of any given time interval (although a caveat is given that the Bouwer result assumes instantaneous sorption and equilibrium---i.e., instantaneous desorption---which implies no effect of sorption at all.) Other textbook treatments similarly provide ostensibly exact paths to \eqref{eq:ADE retarded} without indication that dispersion is being suppressed. \citet[p. 66]{zheng_applied_1995} provide an extensive derivation leading to an apparently exact \eqref{eq:ADE retarded}, but silently introduce an approximation analogous to \eqref{eq:Desorption approximation}, below. \citet[p. 117]{fetter_contaminant_1993} similarly discusses linear isotherms in a transport-free context, and then introduces a retardation constant into the ADE, apparently exactly.
	
	In practice, \eqref{eq:ADE retarded} has also been used for the interpretation of push-pull tracer tests aimed at quantifying $D$ and $R$ \citep{schroth_situ_2000}; ignoring the dispersive effect of sorption. The retarded ADE has also commonly been incorporated in numerical codes that handle more complicated geometries. As the user guide for the popular MT3DMS transport modeling software states, ``{[}i{]}t is generally assumed that equilibrium conditions exist between the aqueous-phase and solid-phase concentrations and that the sorption reaction is fast enough relative to groundwater velocity so that it can be treated as instantaneous....Equilibrium-controlled sorption isotherms are generally incorporated into the transport model through the use of the retardation factor'' \citep[p. 12]{zheng_mt3dms:_1999}. So while it is well established that kinetic mass transfer is a cause of dispersion, the use of retardation factors that ignore it under ``local equilibrium'' conditions is common in practical subsurface hydrology, as well as in the literature. In particular, we note that this is the practice in remediation studies performed on EPA Superfund sites \citep{Chen1999,Zheng1991}, as well as U.S. DOE sites \citep{Rogers1992}. In light of the above, new conceptual arguments pinpointing the approximation being made in the apparently exact derivation of \eqref{eq:ADE retarded} appear timely. 
	
	Regarding the relationship between degree of local equilibrium and usage of the the retarded ADE, there is more to be said. \citet{wallach_small_1998} and \citet{valocchi_validity_1985} acknowledge dispersion due to mass transfer and identify validity of the local equilibrium assumption (LEA) with applicability of the retarded ADE in light of large hydrodynamic dispersion relative to MIMT-driven dispersion (see Appendix \ref{sec: safety}). However, they do not directly investigate the degree of local disequilibrium. By contrast, \citet{bahr_direct_1987} qualify the extent to which fast kinetic mass transfer leads to pointwise local equilibrium (i.e., reduces the difference between $s$ and $(R-1)c$), without directly addressing dispersion due to mass transfer. However, a direct discussion of the degree of support that a given maximum amount of local disequilibrium provides for a given maximum amount of dispersion due to mass transfer (including the potentially surprising answer, \emph{zero}) does not seem to exist in the literature.
	
	In Section 2, we examine mathematically the the implications of the two conceptions of the retardation factor and show how the derivation of the ADE makes a hidden assumption--akin to the gambler's falacy--that hides its inexactitude. In Section 3 we present a numerical study of plume evolution on a heterogeneous 2D conductivity field, as modeled with rapid first-order MIMT and with a retarded ADE. We show a substantial difference in plume evolution despite the fact that local equilibrium is maintained by the mobile and immobile plumes, highlighting the incorrectness of using the term "local equilibrium assumption" to refer to assumed ADE validity. In Section 4, we sum up what we have demonstrated and draw lessons from it. In Appendix \ref{sec: eulerian derivation}, we show how the retarded ADE may be derived as a special case of first-order MIMT in an Eulerian context, and that the remobilization rate is the parameter that controls the divergence between the formulations. In Appendix \ref{sec: safety}, we explicitly discuss past results concerning when it is proper to employ the retarded ADE, highlighting the centrality of the remobilization rate.

\section{Hidden assumptions in the retarded ADE}
	In this section, we establish that interpreting the retarded ADE as exact is to essentially ask for ergodicity to equalize the \emph{absolute} \emph{amount} of time that each particle is immobile in some long time interval, rather than the \emph{fraction} of time immobile. This conflation of absolute and relative frequencies is tantamount to the gambler's fallacy. This fallacy \citep[e.g.,][]{Ayton2004,Sundali2006} represents the erroneous belief that the law of large numbers requires negative auto-correlation in sequences of independent events in order to obtain ``balance'' (informally, that if one has just flipped an unbiased coin for a long string of tails, then heads is now more probable than tails in future flips). In our context, instead of the two states of a coin, we imagine solute particles periodically making a Markovian selection between mobile and immobile states.
	
	It is immediately apparent from viewing $R$ as a scaling factor for time in \eqref{eq:ADE retarded}, that values of $R$ different from unity do not cause any extra dispersion: they simply map the concentration profile at $t$ for any given initial distribution to that at $t/R$ in the case when $R=1$, for the same initial distribution. This is to say: it generates the distribution that would occur if \textit{every} particle spent $t/R$ of the time immobile. If different particles spend different amounts of time immobile during the interval $[0,t]$, then this will represent an additional source of dispersion (which becomes clear when the case $D=0$, $v>0$ in \eqref{eq:ADE retarded} is considered).

\subsection{Relations between mobile and immobile concentrations}
	
	The hypothesis of local equilibrium is \emph{local} in both space
	and time: it constrains the fraction of the solute particles at a
	given location (i.e., small representative pore volume), at any given
	time that are mobile (or equivalently, the instantaneous probability
	that a given individual particle is mobile). The retarded ADE it ostensibly
	justifies depends on a constraint on the \emph{exact amount of time}
	in a given time interval that \emph{each} of the particles is mobile.
	In other words, the retardation factor approach attempts to
	equate an aggregate \emph{spatial} relationship with a deterministic
	temporal quantity. It is important to understand the actual conceptual
	relationship between this spatial constraint and temporal particle
	behavior. To do so we consider the simplest possible mobile-immobile ``transport''
	system---a batch experiment with first-order MIMT---freeing
	us from the need to consider extraneous processes.
	
	Specifically, we consider a steady-state batch system consisting of $N_{m}$ mobile particles and $N_{i}$ immobile particles, where these numbers are both large. For our analysis, we employ the conceptual model implied by system \eqref{eq:Explicit treatment of sorption} which, as we have already mentioned, is a generalization of the retarded ADE, and allows for explicit treatment of individual mobile and immobile intervals.	Our analysis proceeds in a similar spirit to that of \cite{benson_simple_2009}, considering the aggregate behavior that results from independent particles, each of which has the same defined probability distributions for lengths of its mobile and immobile intervals. In this system, the mobile particles have probability $\lambda$ of immobilization per unit time, and the immobile particles have a probability $\mu$ of remobilization per unit time. It follows the expected duration of a single immobilization event is $\mu^{-1}$. We assume all particles are mutually independent and define $K$ as the rate of immobilization, in particles per second: $K=\lambda N_{m}$. Based on the equilibrium conception of retardation and the principle of conservation of mass, the retardation factor satisfies
	\begin{equation}
		R=\frac{\mathrm{E}\left[N_{i}+N_{m}\right]}{\mathrm{E}\left[N_{m}\right]},\label{eq:Retarded}
	\end{equation}
	where $\mathrm{E}[\cdot]$ represents mathematical expectation. Little's law is the intuitive statement that the expected number of
	particles in a state is equal to their rate of arrival multiplied
	by their expected wait in that state \citep[p.  37]{bhat_simple_2008}.
	If the state of interest is the immobile state, this implies $\mathrm{E}[N_{i}]=\mathrm{E}\left[K\right]/\mu$.
	Then we can conclude that
	\begin{equation}
	\begin{array}{rl}
	R & =\frac{\mathrm{E}[N_{m}]\left(\lambda/\mu+1\right)}{\mathrm{E}[N_{m}]}\\
	\\
	 & =1+\lambda/\mu.
	\end{array}
	\label{eq: retardation}
	\end{equation}
	
	It is possible to take this aggregate (multi-particle) spatial behavior
	and draw conclusions about the temporal behavior of any single particle.
	However, we shall see that the validity of these assumptions only
	constrain the \emph{expected} behavior of any particular particle.
	This is to say that if we define $F_{t}$ to be the a random variable
	representing the amount of time a particular particle is mobile in the
	interval $[0,t]$, the assumptions underlying the retardation approach
	will correctly establish that $\mathrm{E}[F_{t}]=t/R$. They will
	not, however, establish that $F_{t}=t/R$, which is what would be
	required for the retarded ADE to be exact. The former condition is
	naturally weaker---constraining only the average of a whole population
	of solute particles---whereas the latter states that \emph{each} solute
	particle in a population is immobile for the same amount of time. It
	is useful to consider these claims precisely.

\subsection{The expected time a single particle is immobile is fixed by $R$}

	By symmetry of particle behavior (i.e., all have the same tendencies
	to immobilize and remobilize), the retardation approach implies that each particle
	is expected (in the mathematical sense) to spend $1/R$ of the \emph{time}
	mobile. To see this, imagine a steady-state, batch system in which $N$
	particles are immobilizing and remobilizing independently of each other. Define,
	for particle $n$, the indicator function
	\begin{equation}
	I_{n}(t)\equiv\begin{cases}
	0 & \textrm{if immobile at \ensuremath{t}}\\
	1 & \textrm{if mobile at }t
	\end{cases},\label{eq:}
	\end{equation}
	which is only non-zero in such cases as the particle is mobile at time $t$. Define 
	\begin{equation}
	\Omega_{t}(N)\equiv\intop_{0}^{t}{\displaystyle \frac{1}{N}\sum_{n=1}^{N}I_{n}(\tau)}d\tau.\label{eq:omega def}
	\end{equation}
	Then 
	\begin{equation}
	\underset{N\rightarrow\infty}{lim}\Omega_{t}(N)=\intop_{0}^{t}\frac{1}{R}d\tau=\frac{t}{R}\label{eq:integration 1st}
	\end{equation}
	which follows, because in the limit $N\rightarrow\infty$, a sample
	mean converges to the expectation (by the law of large numbers), and
	the expected value of an indicator function is the probability of
	being mobile, and $1/R$ of the $N$ particles are mobile at every instant.
	Because of linearity, it is possible to rearrange the order of summation
	and integration, so that
	\begin{equation}
	\begin{array}{rl}
	\underset{N\rightarrow\infty}{lim}\Omega_{t}(N) & =\underset{N\rightarrow\infty}{lim}\frac{1}{N}{\textstyle {\displaystyle \sum_{n=1}^{N}\intop_{0}^{t}{\displaystyle I_{n}(\tau)}d\tau}}\\
	\\
	 & =\mathrm{E}\left[\intop_{0}^{t}{\displaystyle I_{n}(\tau)}d\tau\right]\\
	\\
	 & =\mathrm{E}[F_{t}]
	\end{array}\label{eq:series 1st}
	\end{equation}
	Since $I_{n}(t)$ is just the indicator function that is unity when
	the particle is mobile, the integral represents the amount of time in
	the interval $[0,t]$ in which particle $n$ is mobile. Combining \eqref{eq:integration 1st}
	and \eqref{eq:series 1st}, we see $\mathrm{E}[F_{t}]=t/R$.

\subsection{The absolute time a single particle is immobile is not fixed by $R$}

	The conclusion that $\mathrm{E}[F_{t}]=t/R$ is the strongest that
	can be made. The stronger statement, that $F_{t}=t/R$, for any given
	particle, is false. To see this, consider a system over some interval
	in which all the particles with even index are always immobile and all
	the particles with odd index are never immobile during the interval
	$t$ (not because they are qualitatively different, just that the
	particles are independent and this is one possible, though not likely,
	configuration). Then this system satisfies \eqref{eq:omega def} and
	\eqref{eq:series 1st} for $R=2$, though it is not true for any particle
	that $\frac{t}{2}=\intop_{0}^{t}{\displaystyle I_{n}(t)}dt$.
	
	It is true that, applying the law of large numbers for large $t$,
	it follows that after a long time (i.e., a large number of immobilization and remobilization events), the actual fraction of time every particle
	spends mobile, $F_{t}/t$, approaches $1/R$. Naturally, if each particle
	were to spend \emph{exactly} $1/R$ of the time immobile, for all $t$,
	then we could compute $F_{t}\equiv t/R$ (valid for every particle)
	and the retardation factor approach would be exact. However,
	there is no reason to expect $F_{t}$ to converge to $t/R$ as $t\rightarrow\infty$.
	While possibly unintuitive, such situations are common: consider that
	as $t\rightarrow\infty$, $\left(t+1\right)/t\rightarrow1$ but $\left(t+1\right)\nrightarrow t$.)
	The law of large numbers concerns itself exclusively with \emph{relative}
	frequencies, not absolute frequencies. This is a subtle distinction,
	but an important one: this distinction is what the gambler's
	fallacy (discussed earlier) turns on.

\section{Local equilibrium and retarded ADE validity}

	In this section, we directly consider the degree of support which ``local equilibrium'' (this is to say, fast kinetics) provides to the usage of the retarded ADE. We perform two particle tracking simulations: one employing first-order MIMT, governed by \eqref{eq:Explicit treatment of sorption}, and one employing the retarded ADE \eqref{eq:ADE retarded} with the corresponding $R$ \eqref{eq: retardation}. In so doing, we are able to monitor the degree of local equilibrium between mobile and immobile plumes in the first-order MIMT model, and its coherence with the retarded ADE model that purports to capture it.
	
\begin{figure}
		\centerline{
		\includegraphics[trim={7.5cm 0cm 2.5cm 1cm}, clip, scale=0.75]{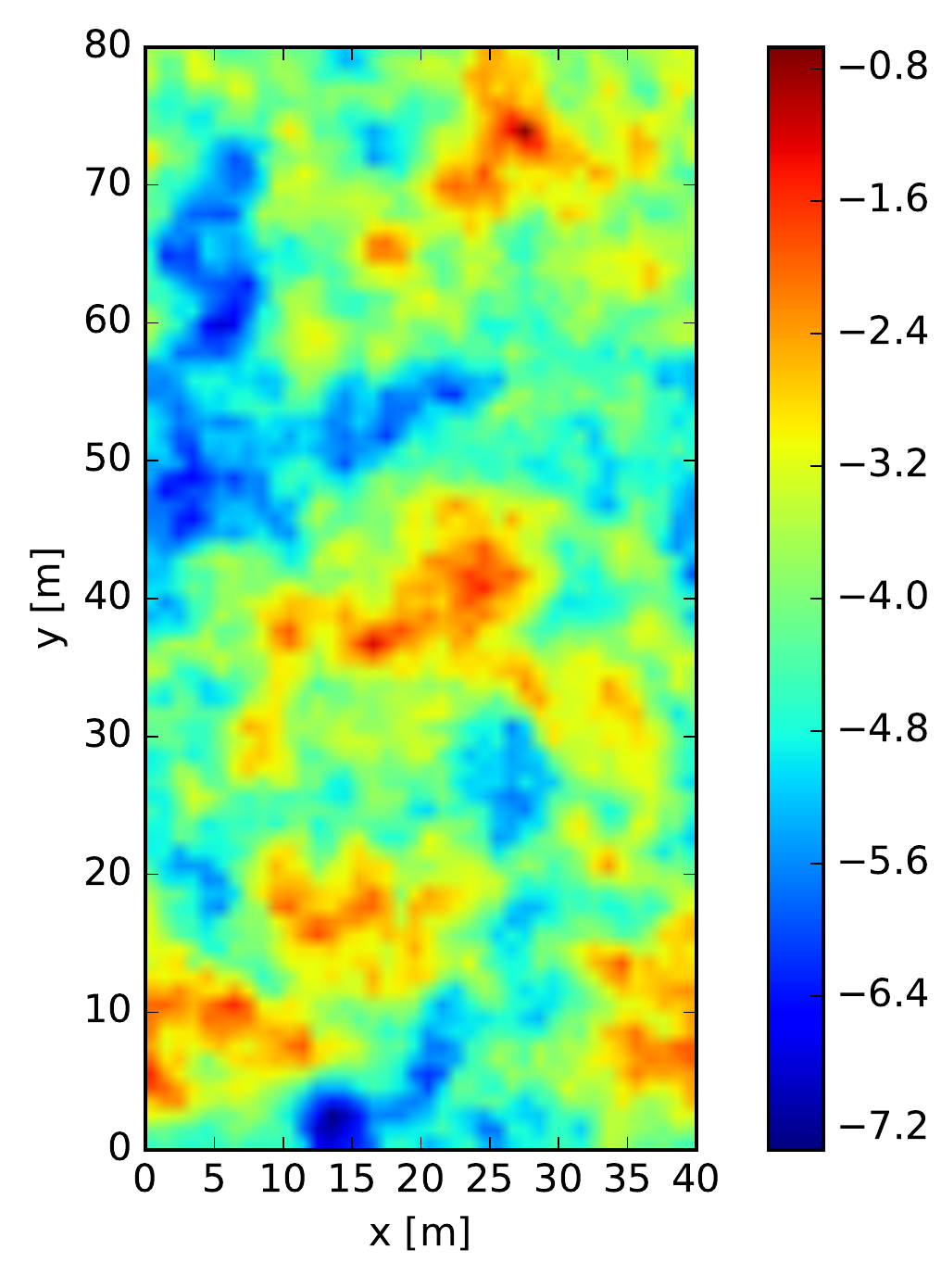}
		\includegraphics[trim={5.5cm 0cm 5.5cm 1cm}, clip, scale=0.75]{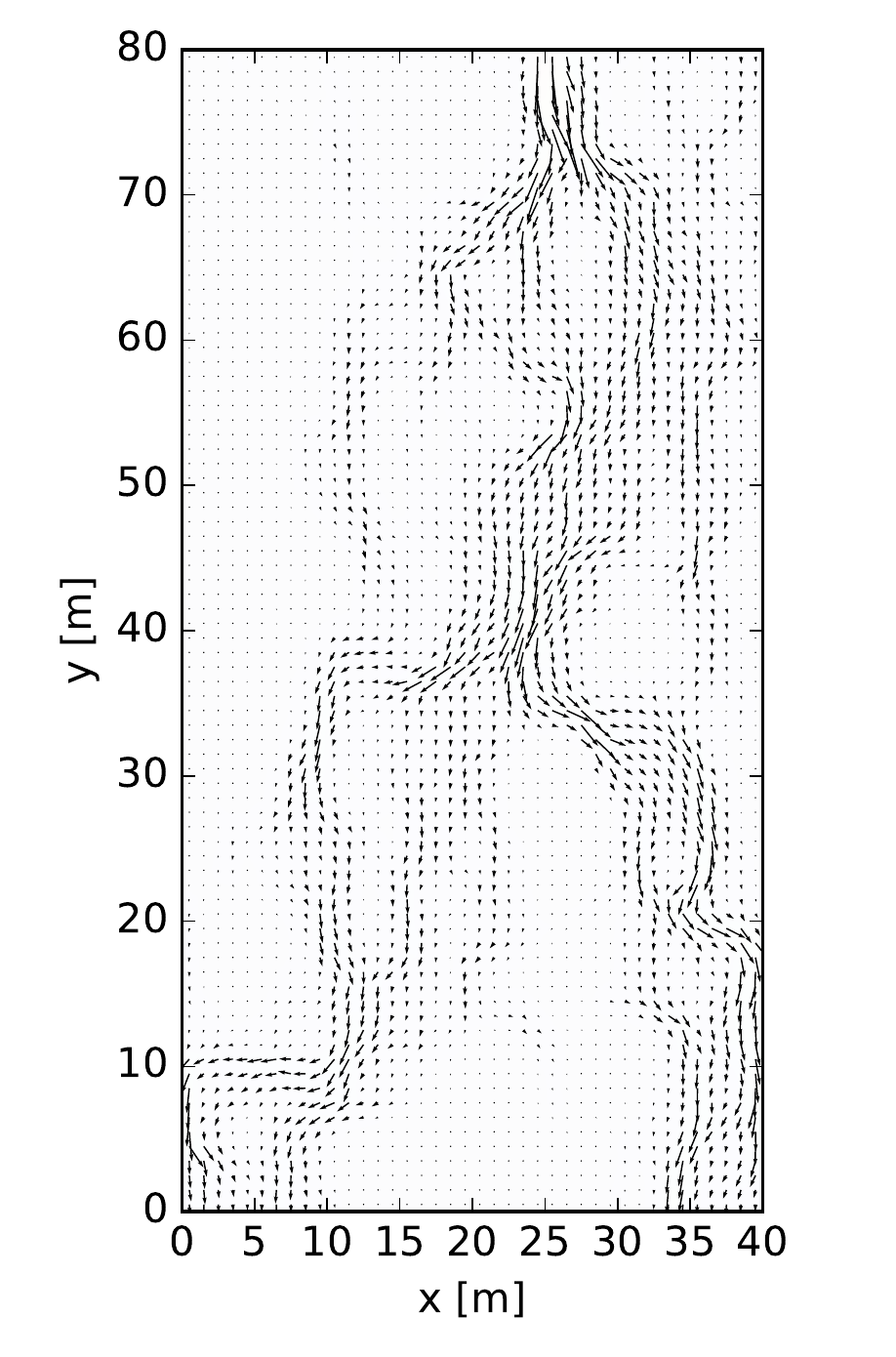}
		}
		\caption{Left: Heat map of $\log_{10} K$, where $K$ is the local hydraulic conductivity of the field in which the particle tracking simulations were performed. Right: quiver plot of heterogeneous velocity field computed by PFLOTRAN using the $K$-field. Each cell-center velocity is indicated by an arrow whose length represents its relative speed and orientation indicates its direction.}
		\label{fig: flow field}
	\end{figure}

	Our study begins by generating a 40 by 80 m random log-hydraulic conductivity field with a multi-Gaussian correlation structure described by an exponential semivariogram with correlation length 5 m, geometric mean conductivity 1e-4 m/s, and $\sigma^2_{\ln K} = 2$ (moderate heterogeneity), discretized into blocks 1 m on a side. The resulting conductivity field is shown in Figure \ref{fig: flow field}. This log-conductivity field is used with the finite-volume numerical flow and transport solver PFLOTRAN \citep{lichtner2015pflotran} to determine the steady-state cell-center velocities. For this computation no-flow boundary conditions are imposed at $x=0$ and $x=40$, a constant pressure of 111135 Pa is imposed at $y=80$, and constant pressure of 101325 Pa is imposed at $y = 0$. The resulting velocity field is also illustrated in Figure \ref{fig: flow field}. Both particle tracking simulations are performed by randomly introducing 500 particles into a circle of radius 2 m, centered at $x=25$ m, $y = 75$ m. When particles are mobile, their positions are tracked by making successive steps of constant duration 0.1 h, during which they passively follow the flow lines. At the end of each step, a small random translation is added to model local-scale dispersion, described by longitudinal dispersivity 0.01 m, and transverse dispersivity 0.001 m. 
	
\begin{figure}
		\centering
	{\fontfamily{phv}\selectfont
		\begin{tabular}{c c c}
			& \parbox{7cm}{\centering\textbf{\textit{t} = 1 y}} & \parbox{7cm}{\centering\textbf{\textit{t} = 5 y}}\\
			\parbox{0.25cm}{\rotatebox{90}{\textbf{Exact first-order MIMT}}}&
			\parbox{7cm}{\includegraphics[trim={5.5cm 0cm 5.5cm 0cm}, clip, scale=0.7]{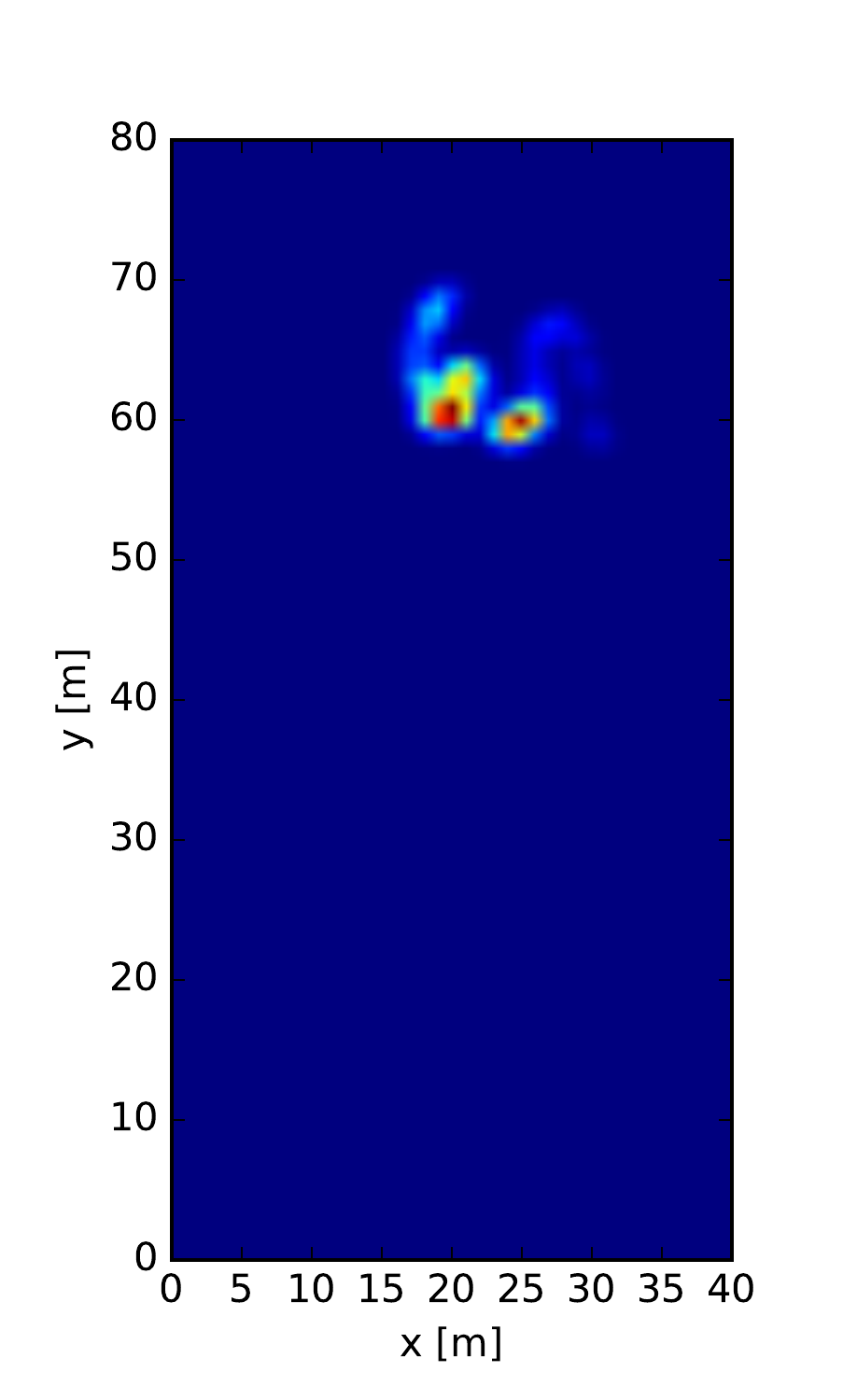}}&
			\parbox{7cm}{\includegraphics[trim={5.5cm 0cm 5.5cm 0cm}, clip, scale=0.7]{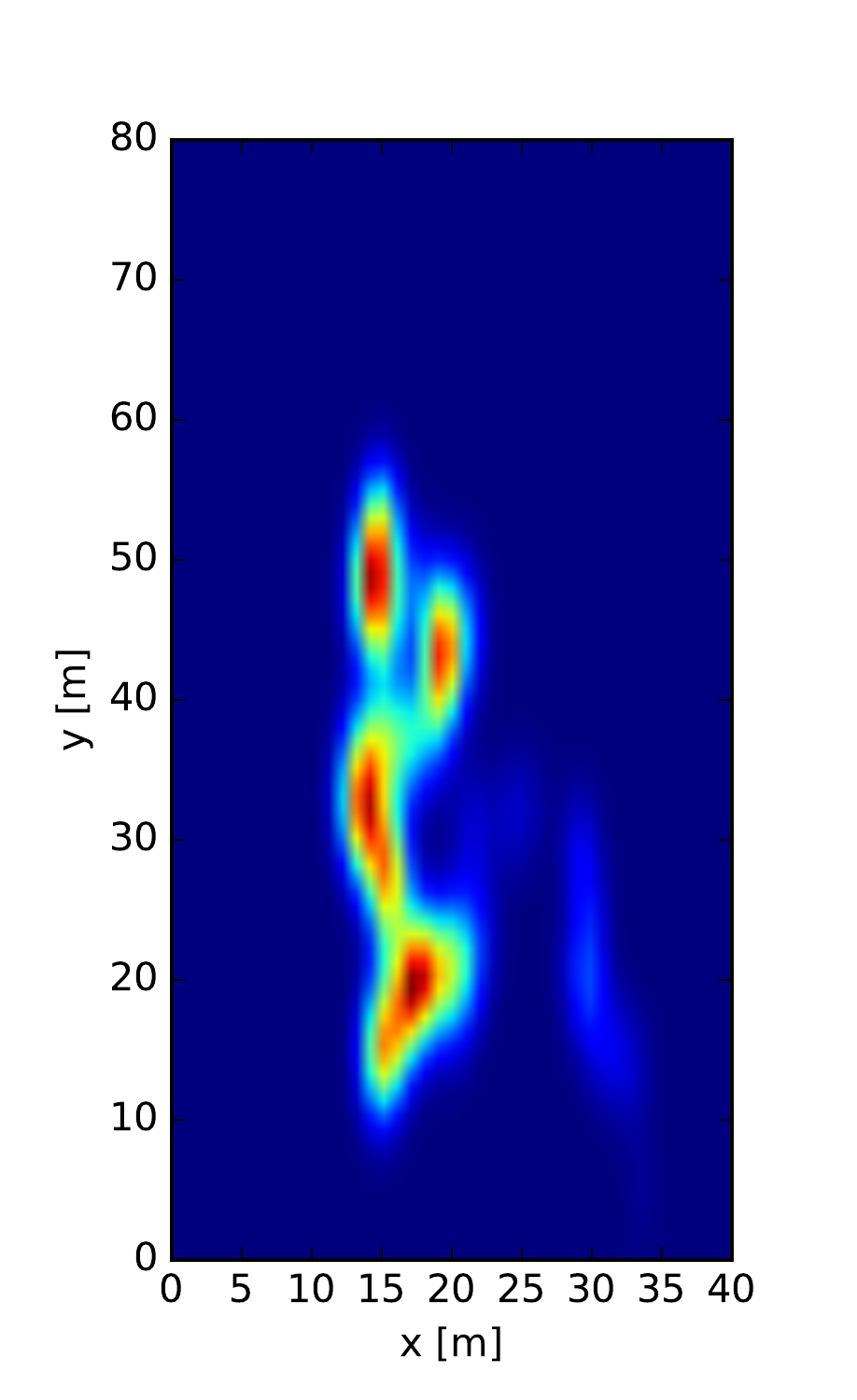}}\\
			
			\parbox{0.25cm}{\rotatebox{90}{\textbf{Retarded ADE approximation}}}&
			\parbox{7cm}{\includegraphics[trim={5.5cm 0cm 5.5cm 1cm}, clip, scale=0.7]{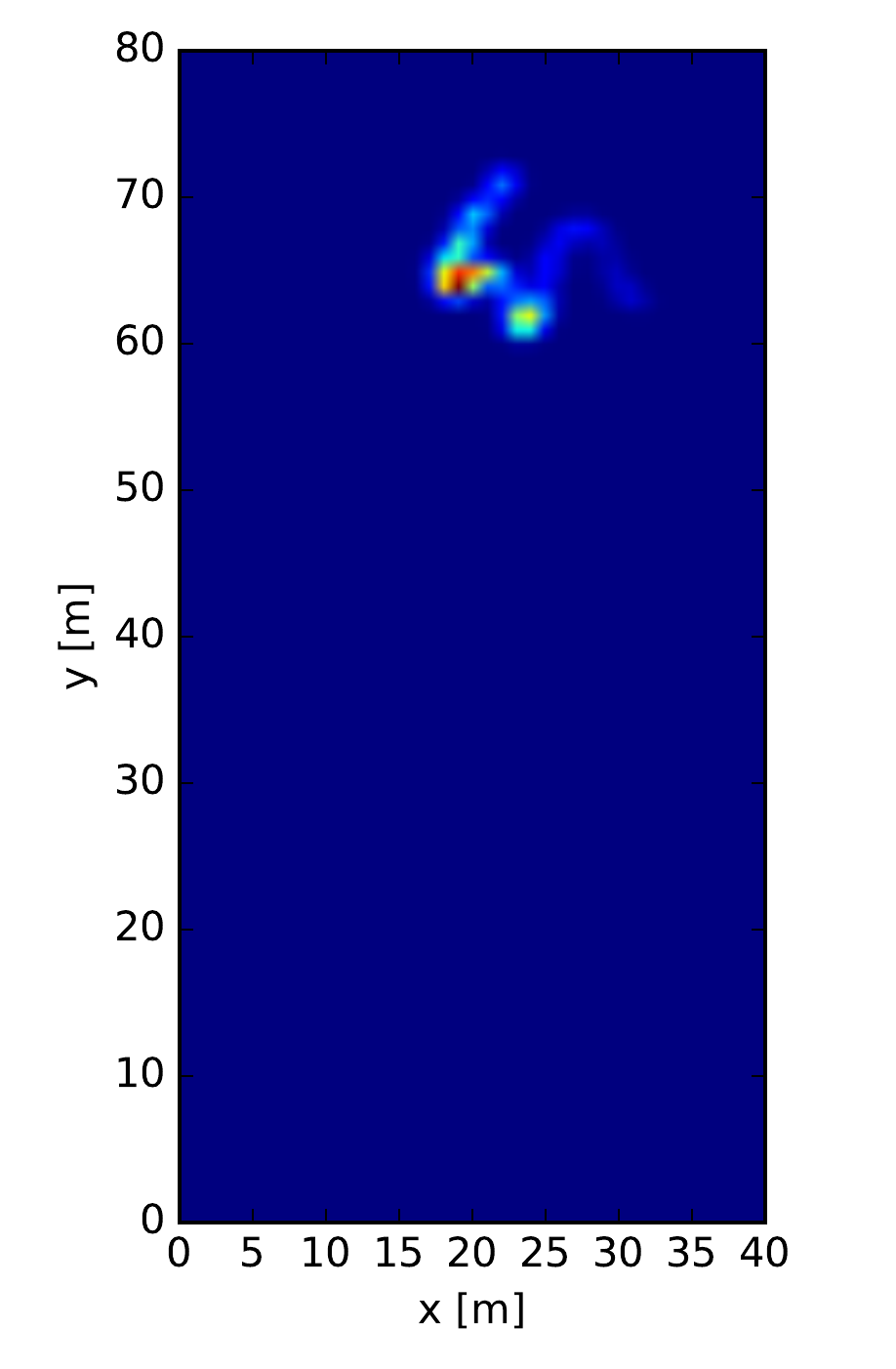}}&
			\parbox{7cm}{\includegraphics[trim={5.5cm 0cm 5.5cm 1cm}, clip, scale=0.7]{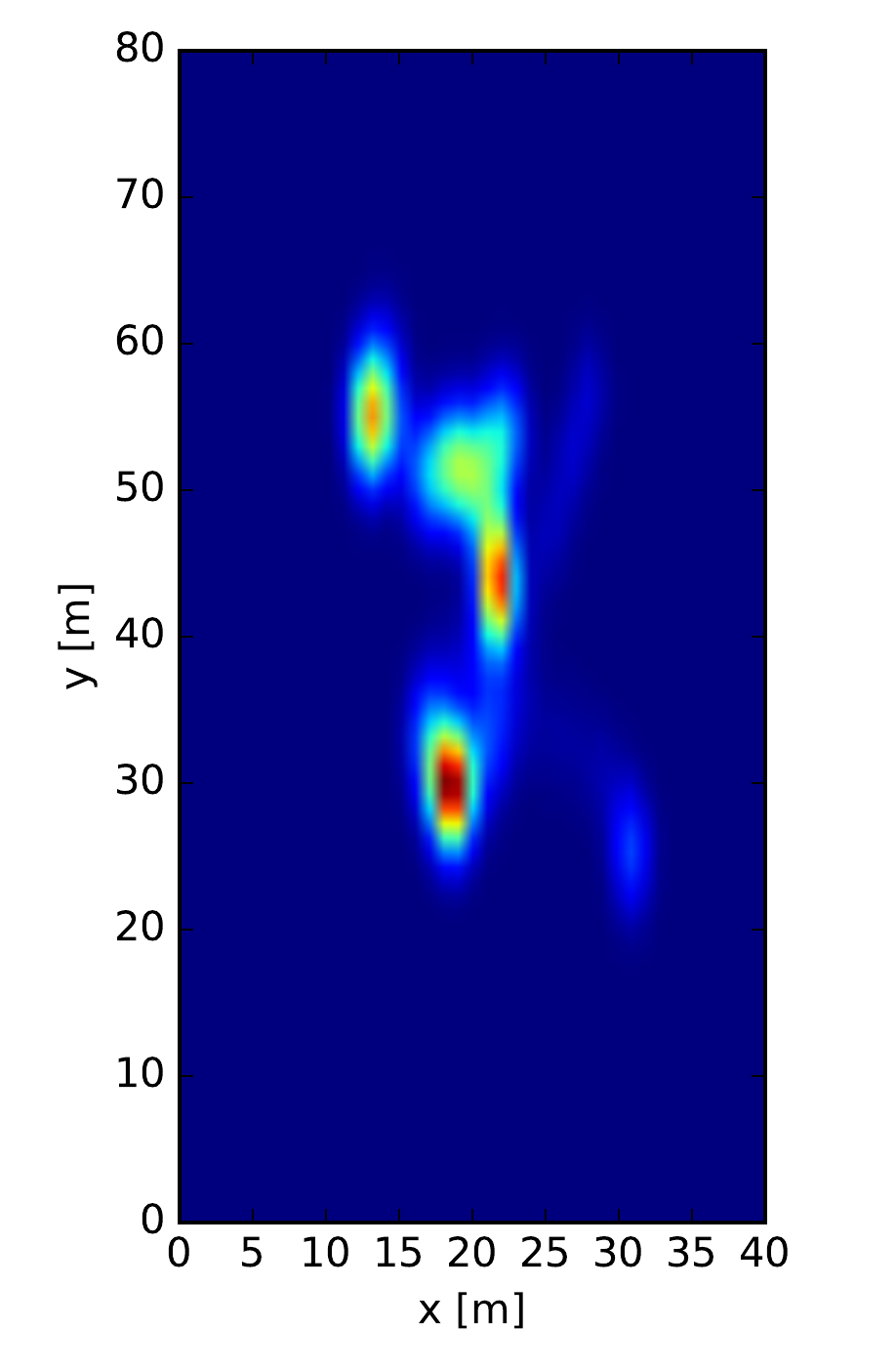}}\\
		\end{tabular}
	}
	\caption{Heat maps of plume concentration at two times under exact first-order MIMT and the retarded ADE approximation.  All plumes used the same velocity field and release location. Hue closer to the red end of the spectrum indicates higher concentration, but scales differ between heat maps.}
	\label{fig:plumes}
	\end{figure}

	For the MIMT simulation, the times of successive immobilization and remobilization events for each particle are generated by draws from exponential random number generators with rate parameters $\lambda=10$ and $\mu=\frac{1}{3}$, respectively. The retarded ADE simulation was performed by disabling particle immobilization altogether and using an alternative velocity field, with identical directions to those used in the MIMT simulation, but all of whose magnitudes were divided by $R$, where $R=1+\frac{\lambda}{\mu} = 31$. Plume concentrations from both particle tracking simulations are determined at $t = 1$ y and $t = 5$ y by performing kernel density estimation using the locations of all particles at the relevant time. These plumes are shown in Figure \ref{fig:plumes}. From examination of the figure, the strong divergence of the two models is apparent.
	
\begin{figure}
		\centering
		\includegraphics[trim={2cm 2cm 2cm 2cm}, clip, scale=0.5]{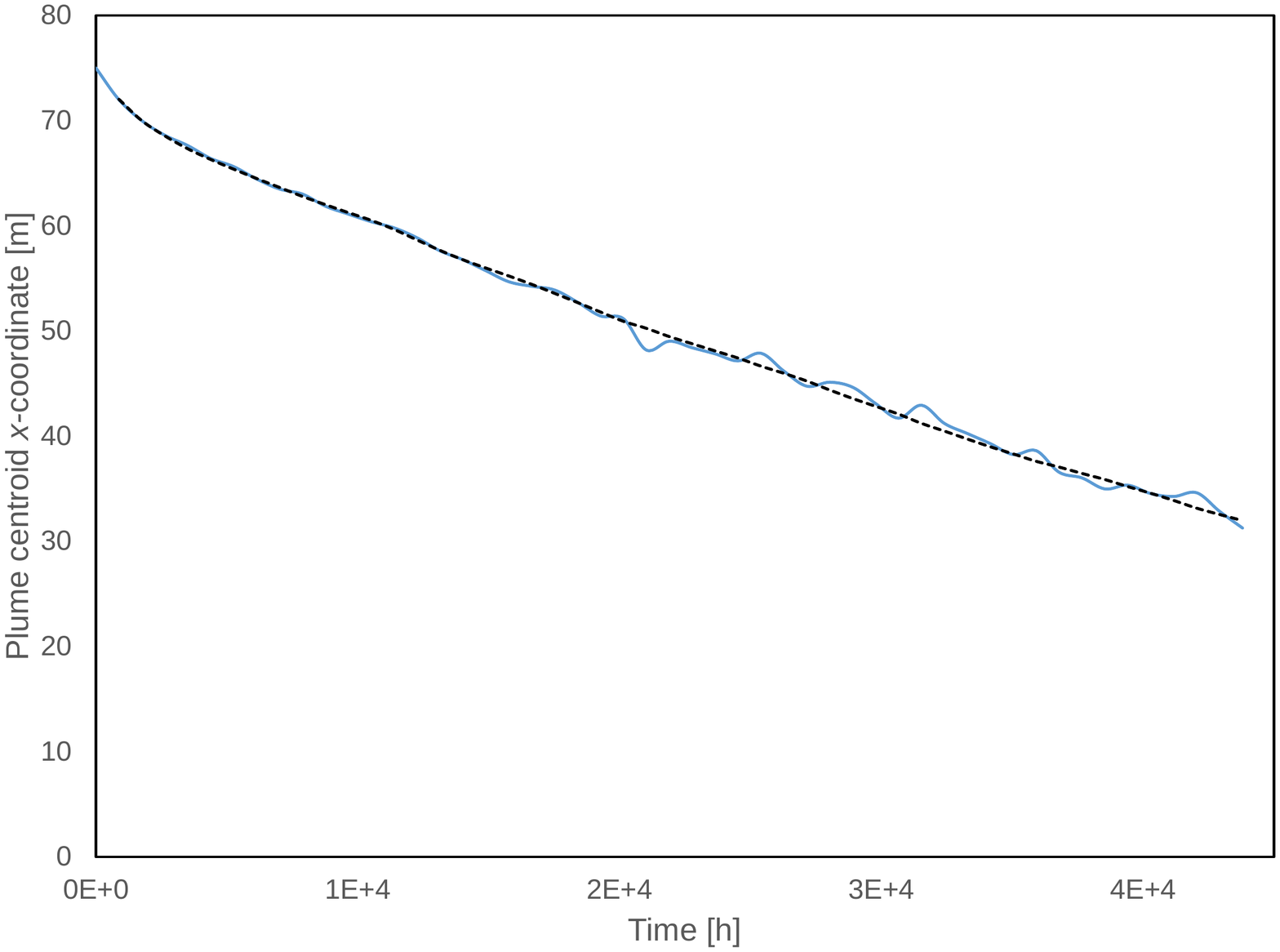}\\
		\includegraphics[trim={2cm 2cm 2cm 2cm}, clip, scale=0.5]{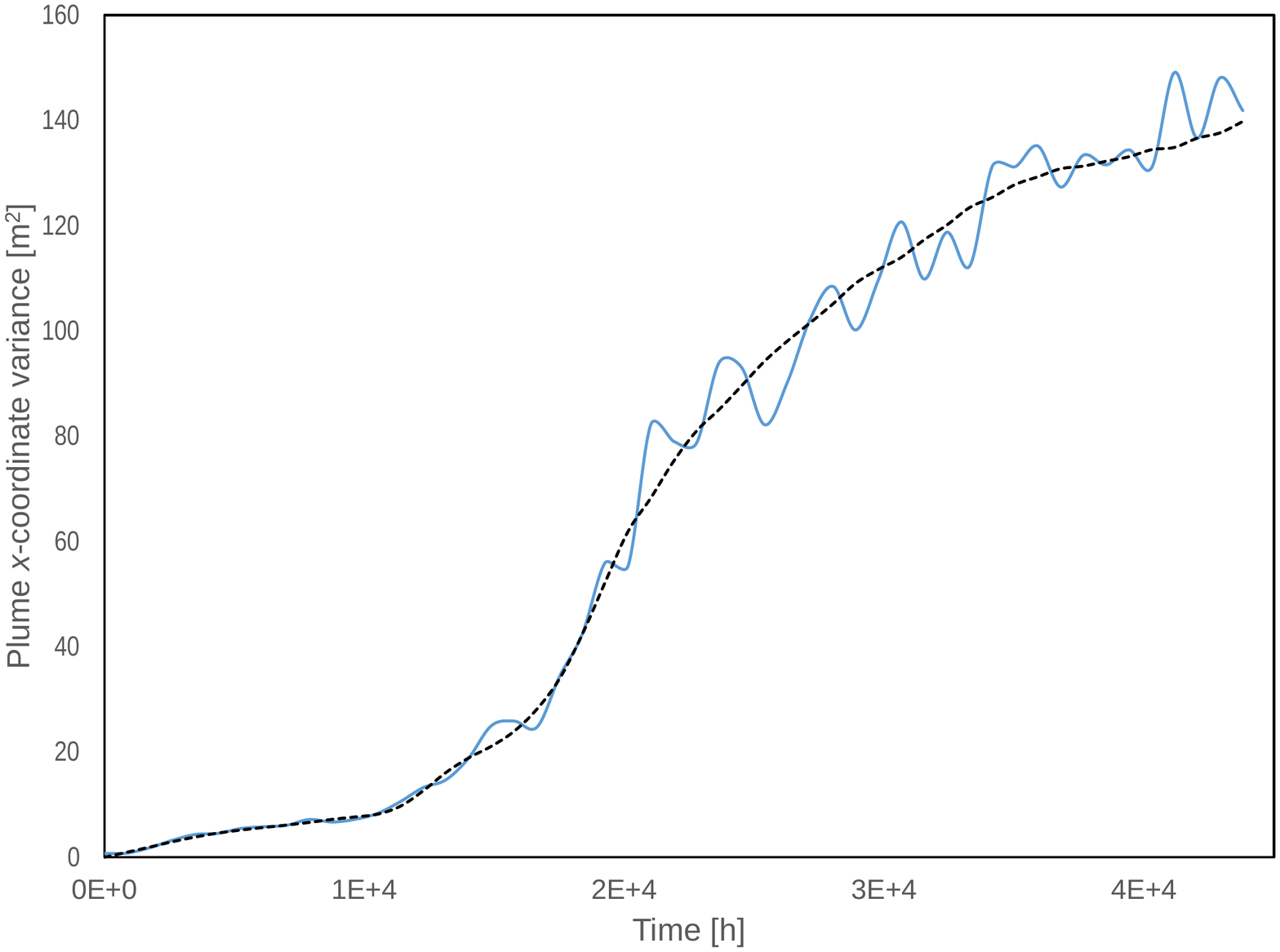}
		\caption{Spatial moments of mobile (solid blue lines) and immobile (dashed black lines) plumes. Top: centroid $x$-coordinate. Bottom: plume $x$-coordinate variance.}
		\label{fig: moments}
	\end{figure}

	Approximate local equilibrium for the MIMT model was established by comparing mobile and immobile plumes at fixed times, and by tabulating each plume's spatial moments over time and verifying their coherence. Graphs of the first two spatial moments are presented in Figure \ref{fig: moments} to illustrate how closely the mobile and immobile plumes cohere. Note that the ratio of immobile to mobile particles is always approximately 30, resulting in smoother immobile particle graphs. We have thus demonstrated an example of a realistic system in which \textit{local equilibrium is satisfied, but performance of the retarded ADE is very poor}. So the use of the term ``local equilibrium assumption'' to refer to the assumption of retarded ADE validity is misleading. Indeed, as we note in Appendix \ref{sec: safety}, classic ``local equilibrium'' metrics actually quantify the relative strengths of MIMT-driven dispersion and hydrodynamic dispersion. They are legitimate metrics for the validity of of the retarded ADE, but do not concern local equilibrium, per se.
	
\section{Summary and concluding discussion}
	The two major contributions of this work are the following:
	\begin{enumerate}
		\item We note that the retarded ADE for \textit{arbitrary} $R$ is derived in canonical sources in a seemingly exact fashion, and is often treated as exact in the literature. We also note that the retarded ADE approach is a special case of first-order kinetic MIMT equations in the limit of instantaneously fast remobilization (i.e. when $R=1$), and otherwise neglects the dispersion that has long been known to be a feature of kinetic mass transfer. We resolve the contradiction by showing how the ostensibly exact derivation of the retarded ADE silently introduces an approximation that is equivalent to the gambler's fallacy.
		
		\item Through a numerical study of transport in a heterogeneous aquifer, we demonstrate that use of the term ``local equilibrium assumption'' to describe the assumption of ADE validity is misleading. In our example, we demonstrate an MIMT-generated plume that is largely disjoint from the plume predicted by use of the corresponding retarded ADE, despite the fact that local equilibrium between mobile and immobile plumes holds everywhere. We note that previously published criteria for the validity of the ``local equilibrium assumption'' are actually criteria for conditions in which the dispersive effect of MIMT is overwhelmed by that of local-scale hydrodynamic fluctuations. We concur that this is the correct condition for usage of the retarded ADE.
	\end{enumerate}
	
	Despite common assumptions to the contrary, the dispersive effect of MIMT, even under ``local equilibrium'' conditions, can not be discarded a priori. Since this extra dispersion may cause un-modeled early- or late-time breakthrough, how to treat it presents a practical question to working hydrogeologists and environmental engineers. The critical role of the remobilization rate, $\mu$, in driving dispersion at late time is clear (see Appendices \ref{sec: eulerian derivation} and \ref{sec: safety}), with only truly \emph{instantaneous} remobilization (i.e., no mass transfer) recovering (\ref{eq:ADE retarded}), and the dispersive effect of sorption increasing as $\mu$ shrinks. Furthermore, MIMT generates anomalous (asymmetric) plumes that are not well described by an ADE at early time.  \cite{hansen_effective_2015} presented the guideline, for small $\mu$, in the $D \rightarrow 0$ limit, that an ADE model becomes adequate after time $t \mu > 70$. Slow mobile-immobile kinetics, and thus and small values of $\mu$, are in reality widespread \citep{pignatello_mechanisms_1996}, so these limitations are practically important.

	We hope that by revisiting this classic topic, we are able to clear up some misconceptions that---as we established in the introduction---continue to persist in the literature, and which have the potential to adversely impact remedial actions.

{
\appendix
\section{Eulerian derivation of retarded ADE from first-order MIMT equations}
\label{sec: eulerian derivation}

	In this paper, we analyze the retarded ADE \eqref{eq:ADE retarded} using the rate constants in \eqref{eq:Explicit treatment of sorption}. To justify this, we show here how \eqref{eq:ADE retarded} represents a special case of \eqref{eq:Explicit treatment of sorption}. The analysis also incidentally shows rapid remobilization as the limiting
	factor for dispersion due to mass transfer. 
	
	We first solve the second of
	equations \eqref{eq:Explicit treatment of sorption} using the standard
	approach for first-order ordinary differential equations to yield
	\begin{equation}
	s\left(x,t\right)=\lambda\intop_{0}^{t}e^{-\mu(t-\tau)}c(x,\tau)d\tau.
	\end{equation}
	Differentiating both sides with respect to time yields
	\begin{equation}
	\frac{\partial s}{\partial t}(x,t)=\lambda c(x,t)-\lambda\intop_{0}^{t}\mu e^{-\mu(t-\tau)}c(x,\tau)d\tau,
	\end{equation}
	and it is apparent from integration by parts that
	\begin{equation}
	\frac{\partial s}{\partial t}(x,t)=\lambda\intop_{0}^{t}e^{-\mu(t-\tau)}\frac{\partial c}{\partial t}(x,\tau)d\tau.\label{eq:Sorbed free convolution}
	\end{equation}
	Note that for large $\mu$ (fast remobilization), $\mu e^{-\mu(t-\tau)}\approx\delta(t-\tau)$,
	the Dirac delta function. This implies that, only in the circumstance of rapid remobilization,
	\begin{equation}
	\frac{\partial s}{\partial t}(x,t)\approx\frac{\lambda}{\mu}\frac{\partial c}{\partial t}(x,t).\label{eq:Desorption approximation}
	\end{equation}
	Substituting this into the first of equations \eqref{eq:Explicit treatment of sorption}
	yields: 
	\begin{equation}
	\left(1+\frac{\lambda}{\mu}\right)\frac{\partial c}{\partial t}(x,t)\approx-v\frac{\partial c}{\partial x}(x,t)+D\frac{\partial c}{\partial x}(x,t)
	\end{equation}
	Thus, employing the large $\mu$ assumption, we can define $R\equiv1+\frac{\lambda}{\mu}$,
	as in \eqref{eq: retardation}, and approximately recover the retarded ADE \eqref{eq:ADE retarded}.
	Note that a memory function convolution such as the one seen in \eqref{eq:Sorbed free convolution}
	generates time ``smearing'' and its neglect when moving to the retarded
	ADE underestimates the resulting dispersion. The relationship used
	to approximately derive the ADE \eqref{eq:Desorption approximation}
	is only exact in the limit as $\mu\rightarrow\infty$, meaning that
	remobilization occurs instantaneously after immobilization. This, naturally,
	generates concentration profiles identical to those in the absence of MIMT.
	
\section{Conditions for proper use of the retarded ADE}
\label{sec: safety}
	The relative effects of local-scale hydrodynamic dispersion, $D$, and dispersion due to MIMT have been explicitly quantified by \cite{Goltz1987} and by \citet{uffink_understanding_2012}, who determined an equivalent effective dispersion coefficient, $D^{\mathrm{e}}$, that describes the behavior of \eqref{eq:Explicit treatment of sorption} at late-time. In our notation:
	\begin{equation} 
	D^{\mathrm{e}}(\lambda,\mu) = \frac{\mu}{\lambda + \mu}D + \frac{\lambda \mu}{(\lambda+\mu)^3}v^{2}.
	\label{eq:Uffink lm} 
	\end{equation}
	Using this expression, (\ref{eq:ADE retarded}) can be recast, using (\ref{eq: retardation}), as:
	\begin{equation} 
	R\frac{\partial c}{\partial t}(x,t)=-v\frac{\partial c}{\partial x}(x,t)+\left(D+\frac{v^{2}}{\mu}\frac{R-1}{R^{2}}\right)\frac{\partial^{2}c}{\partial x^{2}}(x,t).
	\label{eq:Uffink dispersion} 
	\end{equation}
	Clearly, the dispersive effect of MIMT can only be neglected if it is everywhere small relative to the local-scale hydrodynamic dispersion, as encapsulated by $D$. This aligns totally with with the diagnostic criteria for the ``local equilibrium assumption'' derived, using different means, by \citet{wallach_small_1998} and by \citet{valocchi_validity_1985}: $v^{2}/\mu\ll D$. Note that a comparison of dispersive strengths does not quantify local equilibrium!
	
	A point to note---implicit in previous literature, but often neglected in practice---is that one must know the remobilization rate of the MIMT in order to know whether one is making an acceptable approximation in using the retarded ADE. First-order sorption kinetics can be measured in the laboratory under the assumption $\lambda=\mu$ \citep[e.g.,][]{Ho1999,Wu2001,Reddad2002}, and distinct $\lambda$ and $\mu$ can also be measured \citep[e.g.,][]{Meinders1992}. In the case of physical non-equilibrium (i.e., diffusion into secondary porosity), $\mu$ can be approximated from the zero-order terms of the multi-rate mass transfer expressions presented in Table 1 of \cite{Haggerty2000}.
}

\section*{Acknowledgments}

SKH was supported by the LANL Environmental Programs.
VVV was supported by the LANL Environmental Programs and the DiaMonD project (An Integrated Multifaceted Approach to Mathematics at the Interfaces of Data, Models, and Decisions, U.S. Department of Energy Office of Science, Grant \#11145687).

\end{document}